\documentclass[aps,prr,twocolumn,showpacs,superscriptaddress]{revtex4-2}
\usepackage{graphicx}
\usepackage{physics}
\usepackage{float}
\usepackage{bbold}
\usepackage{amssymb}
\usepackage{caption}
\usepackage{subcaption}
\usepackage{comment}
\usepackage{braket}
\usepackage{xcolor}
\usepackage{ragged2e}
\usepackage{hyperref}

\newcommand{\ourrho}{\rho_{\beta}}

\definecolor{DC}{HTML}{28536B}
\definecolor{MDG}{rgb}{0,0.55,0.05}

\definecolor{RefChange}{rgb}{0.0, 0.4, 0.6}  

\begin{document}
\title{Low-Temperature  Gibbs States with Tensor Networks}
\author{Denise Cocchiarella}
\affiliation{Max-Planck-Institut f{\"u}r Quantenoptik, Hans-Kopfermann-Str. 1, D-85748 Garching, Germany}
\affiliation{Munich Center for Quantum Science and Technology (MCQST), Schellingstr. 4, D-80799 M{\"u}nchen, Germany}
\author{Mari Carmen Ba\~nuls}
\affiliation{Max-Planck-Institut f{\"u}r Quantenoptik, Hans-Kopfermann-Str. 1, D-85748 Garching, Germany}
\affiliation{Munich Center for Quantum Science and Technology (MCQST), Schellingstr. 4, D-80799 M{\"u}nchen, Germany}

\date{\today}

\begin{abstract}
 
We introduce a tensor network method for approximating thermal equilibrium states of quantum many-body systems at low temperatures.
Whereas the usual approach starts from infinite temperature and evolves the state in imaginary time (toward lower temperature), our ansatz is constructed from the zero-temperature limit, the ground state, which can be found with a standard tensor network approach.
Motivated by properties of the ground state for conformal field theories, our ansatz is especially well-suited near criticality. Moreover, it allows an efficient computation of thermodynamic quantities and entanglement properties.
We demonstrate 
the performance of
our approach with a tree tensor network ansatz, although it can be extended to other tensor networks, and present results illustrating its effectiveness in capturing the finite-temperature properties in 
one- and two-dimensional scenarios.
In particular, in the critical 1D case we show how the ansatz reproduces the finite temperature scaling of entanglement in a conformal field theory.
\end{abstract}
\maketitle

\section{Introduction}
The Gibbs ensemble describes the state of quantum systems in equilibrium at finite temperature. 
Numerical techniques to approximate these states for quantum many-body systems are thus of fundamental interest. 
Quantum Monte Carlo methods~\cite{sandvik2010computational} provide a powerful approach,  limited, however, to models that do not suffer from the sign problem ~\cite{loh1990sign,troyer2005computational}. 
Tensor networks (TNs)~\cite{Verstraete2008,
Schollwoeck2011,Orus2014annphys,silvi2019tensor,Okunishi2022,Banuls2023} offer a useful alternative, since they are free of sign problem and can efficiently deal with quantum many-body states with moderate correlations. 
In particular, thermal states of local Hamiltonians at any finite temperature satisfy an area law~\cite{wolf2008area} and admit efficient TN approximations~\cite{hastings2006solving,kliesch2014,molnar2015approximating} (see also~\cite{alhambra2023quantum} for more details on the problem of approximating quantum many-body Gibbs states).

Regarding practical methods, the most common TN algorithms for finite temperature
try to find thermal states via
imaginary time evolution ~\cite{verstraete2004matrix,zwolak2004,feiguin2005finite}.
Namely, starting from a maximally entangled state between the system and an ancillary copy (which can be understood as a maximal-rank thermal ensemble), the application of the imaginary time evolution operator prepares a purification of the Gibbs ensemble (the thermofield state).
This method
usually works well at high temperatures, but accumulates error while evolving to low-temperatures, although several strategies have been developed to mitigate this ~\cite{chen2018exponential,Li2023tangent}. 
Another drawback
of the thermofield
strategy is that
the rank of the 
state is never decreased in the thermal evolution, even though at zero temperature the state should reduce to the rank-one density operator of the ground state (assuming its uniqueness).
A few TN methods exist
that use a low-rank description.
Samples of minimally entangled typical thermal states (METTSs)~\cite{white2009minimally,stoudenmire2010minimally,sinha2024efficient}, have proven to be effective in representing thermal properties, especially in gapped systems at temperatures well below the energy gap~\cite{binder2015minimally}. 
In Ref.~\cite{Iwaki2021tpq-mps}, a single matrix product state (MPS) was constructed to represent a typical thermal pure state~\cite{Sugiura2012tpq}, and recover physical quantities with almost no sampling cost.
A recent proposal used tree tensor operators (TTOs)~\cite{arceci2022entanglement} to represent
low-temperature thermal states from a few individually approximated excitations. An intermediate approach is taken in~\cite{reinic2024finite}, where the standard purification algorithm is run, but then the resulting state is approximated with a low-rank TTO.

Here we propose a method that directly targets the low energy regime to construct a low-rank TN approximation starting from the zero-temperature limit, and thus is complementary to the standard imaginary time evolution.
As the few-excitations ansatz using TTOs~\cite{arceci2022entanglement}, it allows one to efficiently compute entropy and entanglement properties of mixed states, but it is computationally more efficient. 

Our method is inspired by 
the properties of
critical systems described by a conformal field theory (CFT)~\cite{francesco2012conformal}. 

In a CFT, the ground state
 plays a key role in the creation of the conformal tower, acting as the primary state to generate excited states through specific local operators ~\cite{francesco2012conformal,herwerth2015excited}. Similarly, we start from the ground state to define a relevant subspace for the approximation of a thermal state. 
Thus our method is especially well suited for systems near criticality.
 We nevertheless show how the applicability of the method actually extends to more general low-temperature scenarios and check this with numerical examples in
one- and two-dimensional models.
The rest of the paper is organized as follows. In Sec. \ref{sec:method} we present the method, starting with the construction of the Gibbs state from the local effective Hamiltonian, followed by the implementation details using Matrix Product States (MPS) and Tree Tensor Networks (TTN), and concluding with an efficient approach for computing thermal and entanglement properties. Section \ref{sec:numerical} provides numerical results, including applications in both one-dimensional and two-dimensional systems. Since the approach performs particularly well near criticality, Sec. \ref{sec:non-crit}  is devoted to improving its accuracy and applicability away from critical points.
Finally, Sec. \ref{sec:discussion} offers a broader discussion of the results and potential directions for future work.

\begin{figure}[h]
    \centering
    \includegraphics[width=1\linewidth]{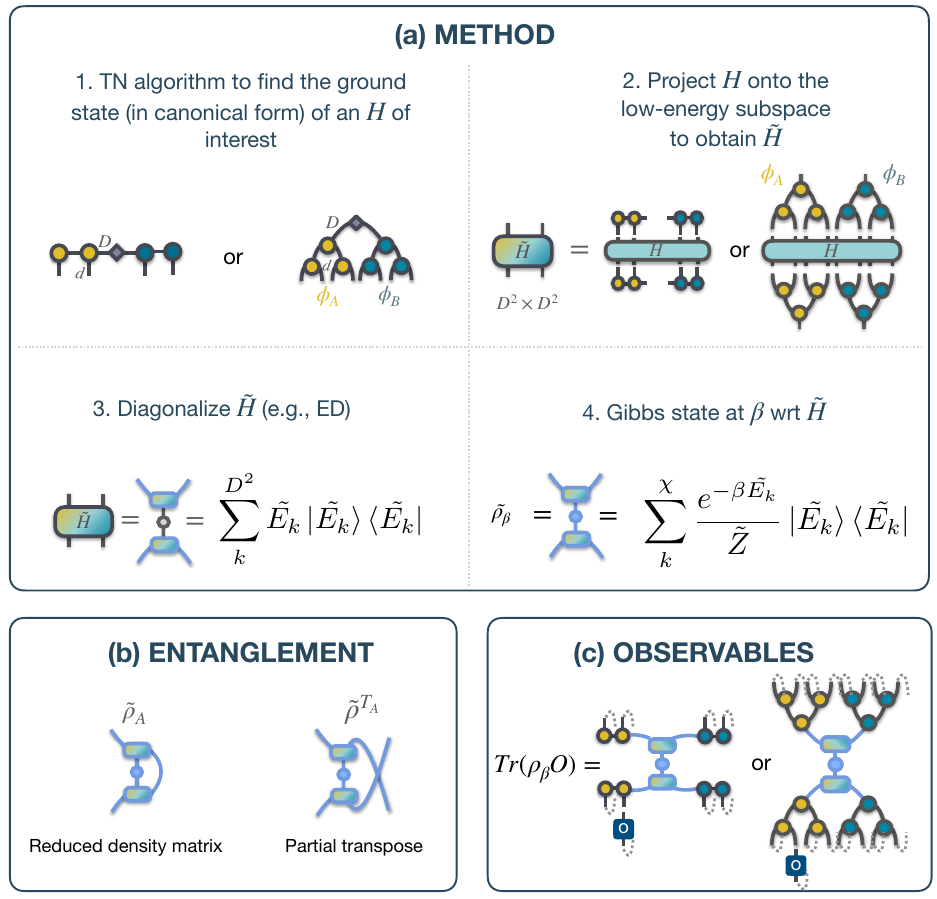}
    \caption{ \justifying Graphical representation of the proposed TN method. (a) Description of the method. 1. The TN for the ground state in canonical form with respect to a bond defines the isometries $\Phi_{A(B)}$; $d$ is the physical dimension and $D$ is the bond dimension. 2. Projecting the full Hamiltonian with the Schmidt vectors of the ground state $\Phi_{A(B)}$ yields the corresponding effective Hamiltonian $\tilde{H}$. The dimension of the isometry $\phi_A \otimes \phi_B$ is $D^2 \times 2^N$, yielding an effective Hamiltonian $\tilde{H}$ dimension of $D^2 \times D^2$. 3. Diagonalizing $\tilde{H}$, through e.g., exact diagonalization (ED), provides the set of eigenvalues $\tilde{E}$ and eigenvectors $\ket{\tilde{E}}$. 4. Thermal state representation via TN of the reduced Hamiltonian $\tilde{H}$ and its decomposition in terms of energy eigenvalues and eigenvectors.
    The circle in the vertical bond represents the Boltzmann weights from $\tilde{E}$. (b) Basic elements for the computation of entanglement quantities. On the left, a TN representation of the reduced density matrix when subsystem B is traced out, $\tilde{\rho_A}$. This would allow computing the Von Neumann entropy of the thermal state. On the right, a TN representation of the partial transpose $\tilde{\rho^{T_A}}$, necessary when computing logarithmic negativity, a valid measure of entanglement for mixed states \cite{plenio2005logarithmic}.
    (c) Basic elements for the computations of observables and expectation values. First, $\tilde{\rho}$ needs to be mapped back into the physical space, using the isometries from step 1(a), yielding an approximation for the thermal state $\rho_\beta$. Then, any operator $O$ can be applied to the TTO. In picture, a single site operator, acting on site 2, $O_2$. To compute $\Tr(\rho_\beta O)$, we need to perform standard contractions of the physical legs of the resulting TTO, as graphically indicated by the dashed lines. 
    }
    \label{fig:method}
\end{figure}

\section{METHOD}
\label{sec:method}

\subsection{Gibbs state from the local effective Hamiltonian} 
\label{subsec:gibbs}

At finite temperature $T=1/\beta$, the equilibrium state of a system governed by a Hamiltonian $H$ is described by the Gibbs ensemble
$\rho_{\beta}=e^{-\beta H}/Z=\left(\sum_{n}e^{-\beta E_n} \ket{E_n}\bra{E_n}\right)/Z$, 
where $\ket{E_n}$ ($n=0,\ldots d^{N}-1$) are the 
eigenstates of $H$
with energies $E_n$, and $Z=\text{Tr}(e^{-\beta H})$ is the partition function. 
At zero temperature ($\beta\to\infty$), and assuming the ground state $\ket{E_0}$ is unique, this reduces to the rank-1 density operator $\rho_{\infty} =\ket{E_0}\bra{E_0}$. 
At finite $\beta$, contributions from higher energy states are suppressed exponentially by their excess energy above the ground state, and thus at low temperatures we may 
expect the Gibbs state to be well approximated by a low-rank operator.

Ref.~\cite{arceci2022entanglement} proposed a TN algorithm to explicitly construct such a low-rank approximation
by  approximating each of the few first excitations by a tree tensor network (TTN) ansatz, found variationally, and explicitly constructing their mixture with corresponding Boltzmann weights.
This approach can accurately capture the behavior at very low temperatures, but becomes computationally challenging when the number of
required levels 
increases, for instance in cases with closely spaced energies, as in critical systems. Moreover, the mixture constructed using a given number of eigenstates does not necessarily produce the best approximation with fixed rank.

It is thus more desirable to have an efficient description of a relevant subspace that allows us to reconstruct the low-temperature properties.
In the case of critical systems described by a CFT, 
such subspace could be spanned by the Schmidt vectors of the ground state, based on the observation that excited states can be expanded in this basis~\cite{cocchiarella2025excited}.

This suggests a method to obtain the low-rank approximation of $\rho_{\beta}$ from the ground state of a 1D critical system.
Considering a 
bipartition of the ground state
into subsystems $A$ and $B$, the corresponding Schmidt decomposition reads
$\ket{\psi_{GS}} = \sum _{\alpha}^{D} \lambda_{\alpha} \ket{\phi_A^{\alpha}}\ket{\phi_B^{\alpha}}$, where 
$\ket{\phi_{A(B)}^{\alpha}}$ are the Schmidt vectors for the $A$ ($B$) partition.
They define isometries $\Phi_{A(B)}$ that map the corresponding $d^{N_A}$ (resp. $d^{N_B}$) dimensional physical space to a virtual degree of freedom of dimension $D$.
Their tensor product $U=\Phi_A \otimes \Phi_B$ is thus an isometry that maps the Hamiltonian onto the $D^2$-dimensional subspace spanned by tensor products of Schmidt vectors, as $\tilde{H} = U HU^\dagger$. For a given inverse temperature $\beta$, we can then approximate the Gibbs state by that of the effective Hamiltonian $\tilde{H}$, embedded in the physical space as $e^{-\beta H} \approx e^{-\beta U^\dagger \tilde{H} U}= U^\dagger e^{-\beta U H U^\dagger} U$.

\subsection{Implementation with MPS/TTN}
 \label{subsec:mps-ttn}
 
 This scheme is readily implementable with MPS and TTN with the following algorithm, which we illustrate graphically in Figure \ref{fig:method}.
 \begin{enumerate}
     \item 
Given the Hamiltonian of interest H, we start by variationally optimizing the corresponding TN 
representation of the ground state 
with bond dimension $D$, 
and write it in canonical form with respect to a chosen bipartition (e.g. illustrated is the half system case).
The Schmidt vectors $\ket{\phi_{A(B)}^{\alpha}}$ then directly correspond to each half-network, as indicated in the figure by yellow and blue colored tensors.
     \item
 The original Hamiltonian is then projected onto the space spanned by the products of Schmidt vectors. 
     \item
Diagonalizing the projected $\tilde{H}$ yields a set of eigenvalues $\{\Tilde{E}_k\}$ and eigenvectors $\{\ket{\Tilde{E}_k}\}$ where $k=1,...,\chi = D^2$. 
This diagonalization can be done exactly, as we explore in this work, or, for larger values of $D$, using iterative methods that target a fraction of the eigenstates.
     \item
     The thermal state of the reduced Hamiltonian $\tilde{H}$
     for inverse temperature $\beta$, i.e. the state that minimizes the free energy, is constructed as
     $\tilde{\rho}=\left(\sum_{k=0}^{\chi} e^{-\beta\Tilde{E_k}}\ket{ \Tilde{E_k}}\bra{\Tilde{ E_k}}\right)/\tilde{Z}$, 
     with $\tilde{Z}=\sum_{k'}e^{-\beta \Tilde{E_{k'}}}$. 
 \end{enumerate}

Our ansazt for the thermal state of $H$
is then obtained by mapping back $\tilde{\rho}$ onto the physical space, as graphically shown in Fig.\ref{fig:method}c), allowing to compute observables and expectation values.
Explicitly,
 \begin{equation}\label{thermalstate}
     \ourrho = U^\dagger \sum_{k=0}^{\chi} \frac{e^{-\beta\Tilde{E_k}}}{\tilde{Z}}\ket{ \Tilde{E_k}}\bra{\Tilde{ E_k}}U=  U^\dagger \tilde{\rho}U.
 \end{equation}
 
Notice that the reduced Hamiltonian $\tilde{H}$ corresponds to the effective Hamiltonian on a bond of the tensor network.
This construction is closely connected to the constructions 
in~\cite{chepiga2017excitation,eberharter2023extracting}, in which the effective Hamiltonian corresponding to one- or two-
sites was extracted from a MPS approximation to the ground state, and shown to access a significant part of the spectrum of the original Hamiltonian in the critical case. 
Our Hamiltonian, instead, is the effective one for \textit{a zero-site} block, and thus 
corresponds to the effective Hamiltonian in the subspace generated by 
tensor products of
the ground state Schmidt vectors.

\begin{figure*}[htpb]
    \centering
     \includegraphics[width=1\textwidth]{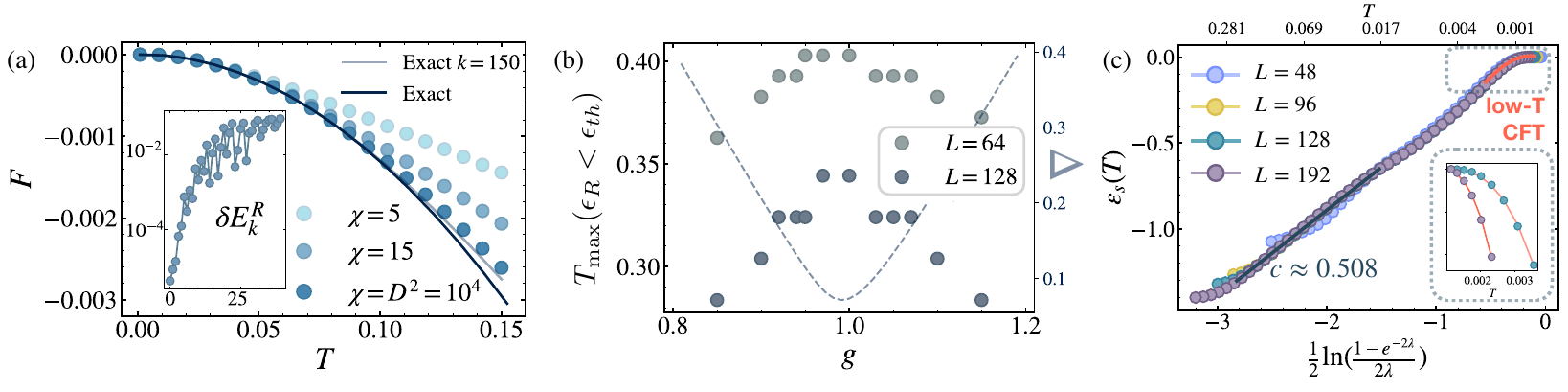}
    \caption{ \justifying Numerical results for the $1\mathrm{D}$ TFI. 
    (a) Free entropy {(main plot)} 
    and relative error in energy spectrum $\delta E_k^R$ (inset)
    at $g=1$, $L=128$ with open boundary conditions. Error bars (obtained comparing $D'=70$ and $D=100$) are smaller than the size of the markers.
    (b) Dependence on the field $g$ of the maximal temperature
    that can be simulated while keeping the relative error in free energy
    below
     a predefine threshold $\varepsilon_{th}=10^{-2}$. Two system sizes $L=64,\ 128$ are shown (filled symbols), along with the exact energy gap $\Delta$ computed for $L=128$ (dashed line).
    (c) Entanglement negativity for the critical case with 
    periodic boundary conditions. 
    The inset shows a fit of our data to the low-temperature prediction of $\varepsilon_s(T)$. The bond dimension used for all system sizes is $D=10^2$ and the number of excited states included in the thermal ensemble is $\chi=10^4$.}
    \label{fig:1d}
\end{figure*}

\subsection{Efficient calculation of thermal and entanglement properties} 
\label{subsec:efficient}

With this construction, the thermal state is described by a low-rank mixed TN
(illustrated in Fig.\ref{fig:method} (c)). 
Expectation values of few-body observables can be efficiently computed in this state, using standard MPS~\cite{Verstraete2008,Schollwoeck2011} or TTN~\cite{shi2006classical,silvi2019tensor} contractions.

Thermodynamic quantities can be easily obtained from the spectrum of $\tilde{H}$. In particular, the free energy reads
\begin{align}
F(\ourrho) &= \Tr(H \ourrho)-{T}S(\ourrho) \nonumber \\
&= \Tr(\tilde{H} \tilde{\rho})-{T}S(\tilde{\rho})=-T\ln \tilde{Z}, 
\label{eq:freeE}
\end{align}
using the cyclic property of the trace and the fact that $\rho_{\beta}$ and $\tilde{\rho}$ are related by an isometry. All the other thermodynamic properties can be then obtained starting from the free energy.

The ansatz allows also access to entanglement quantities, typically not accessible from the most standard purification algorithm. 
The isometry relating $\rho_\beta$ to $\tilde{\rho}$ has a tensor product structure, therefore it does not change the entanglement properties with respect to the corresponding bipartition.
It is then possible to compute such quantities directly for $\tilde{\rho}$, without the need for mapping the state back onto the physical space.

As an example, we describe
how to compute the logarithmic negativity \cite{plenio2005logarithmic,peres1996separability} (see the right side of Fig~\ref{fig:method} (b)). 

Let us consider two partitions $A$ and $B$ described by a 
density matrix $\rho_{AB}$, 
and denote by $\rho^{T_A}$ the partial transpose 
with respect to partition $A$. The logarithmic negativity is defined as  
\begin{equation}
\varepsilon(\rho)=\ln{(2\mathcal{N(\rho)}+1)},
\label{eq:logN}
\end{equation}
where the negativity $\mathcal{N(\rho)}$ 
is the sum of negative eigenvalues of $\rho^{T_A}$,
\begin{equation}
\mathcal{N(\rho)} = \sum_{ \sigma(\rho^{T_A}) < 0} \abs{\sigma(\rho^{T_A})}.
\label{eq:negativity}
\end{equation}
$\mathcal{N(\rho)}$ is positive if and only if 
the partial transpose of $\rho$ is not positive, i.e.
$\rho$ violates the positive partial transpose (PPT) condition~\cite{peres1996separability,horodecki1996information,horodecki2009quantum,vidal2002computable}. While being a valid entanglement measure for mixed states, computing the negativity requires the full spectrum of $\rho^{T_A}$, 
which in general is exponentially costly in a many-body state.
With our ansatz, however, the spectrum of $\rho^{T_A}$ is equivalent to the spectrum of the partial transpose of $\tilde{\rho}$, $\tilde{\rho^{T_A}}$, which can be computed in the reduced space (see also~\cite{arceci2022entanglement,reinic2024finite}).

\section{Numerical results}
\label{sec:numerical}
In this section, we benchmark the performance of the method for one- and two-dimensional problems, using a TTN ansatz. In the one-dimensional case we
find that, as can be expected, at, and near, critical points (i.e., when the energy gap is relatively small), the method works at its best. 
When the gap increases, the performance worsens, reflecting the fact that the ground state Schmidt vectors in this case do not capture a
large enough subspace~\cite{chen2018exponential}. We find that the method can be improved by combining the information of further excitations above the ground state, and using the effective Hamiltonian for the highest computed one~\ref{sec:non-crit}, we can thus interpolate between our method and the one in Ref.~\cite{arceci2022entanglement}. Finally, we apply the method to a two-dimensional system ~\ref{subsec:num2D}. 

\subsection{Numerical results in 1D}
\label{subsec:num1D}

To benchmark the method, we focus on
the one-dimensional transverse field Ising ($1\mathrm{D}$ TFI) model, whose Hamiltonian reads
\begin{equation}
    H = J \sum_{\langle i,j\rangle} \sigma_i^z \sigma_j^{z} - g\sum_i\sigma_i^x,\label{hising}
\end{equation}
where $\langle i,j\rangle$ denotes the summation over nearest-neighbor pairs.
The analytically solvable 1D TFI has
a phase transition at $g_c/J=1$, and thus 
allows us to benchmark our results in different regimes (in the rest of the paper, we set $J=1$).
First, we compare the free energy obtained with our method, using a TTN ansatz, with the exact value. 
The results, in Fig.~\ref{fig:1d}(a) (dark blue symbols), show that the ansatz correctly reproduces a range of temperatures $T\lesssim 0.1$, before deviating from the exact value (solid black line). The results shown in the plot correspond to bond dimension $D=100$, enough for convergence, as indicated by the error bars, smaller than the symbols (see Appendix \ref{app:convergence} for details on the numerical errors).

In order to investigate the source of the error, we compare our results with those from the exact solution truncated to a certain number of excitations.
We observe that 
the accuracy of our result is
comparable to including the first $k=150$ exact excitations in the mixture (blue solid line).
We may wonder whether this corresponds to the number of individual eigenenergies that get accurately captured by $\tilde{H}$.
However, 
this is not the case, as shown in
the inset of Fig.~\ref{fig:1d},
which displays the
relative error in the successive eigenvalues of $\tilde{H}$ with respect to the exact energies. This error
is already on the order of $10^{-1}$ after $\sim 20$ energy levels. 

Nevertheless, it is worth noticing that the first $10-12$ excitations are obtained with high accuracy, with a 
relative error below $1\%$, and this at a 
low computational cost: our method requires a single run of the variational tensor network optimization, i.e., a single state, to obtain the effective Hamiltonian.
This is to be compared with
the usual strategy to obtain multiple excitations with a TN ansatz, which would  require the sequential optimization of each level, with overall cost scaling with the square of the number of desired excitations \cite{McCulloch_2007,schollwock2005density}(in particular, for trees~\cite{silvi2019tensor} $\mathcal{O}(k^2 D^4)$).

Our current implementation of the method fully diagonalizes $\tilde{H}$, with a computational cost $\mathcal{O}(D^6)$ (see Appendix \ref{app:cost} for details).
Whereas the rank of the effective Hamiltonian can be $D^2$ (see Fig.\ref{fig:method}), we can further truncate
the effective Gibbs ensemble by keeping a smaller number of effective eigenstates $\chi$ (truncating the sum in Eq.\ref{thermalstate}), so that we can mimic the result of a few-excitation ansatz.
The lighter symbols in Fig.~\ref{fig:1d}(a) show
 how 
the higher the value of $\chi$, the larger the range of temperatures that are accurately described.
Because $1\leq \chi \leq D^2$, extending the temperature range
even more would require increasing the bond dimension $D$ of the ground state, which serves as a free parameter in our method. Increasing $D$ expands the Hilbert space accessible to the Schmidt vectors, with the dimension scaling as $D^2 \times D^2$. 

Even though the method is inspired by CFT arguments, we can 
probe the ansatz away from criticality.
To evaluate performance in such cases, we estimate the maximum temperature $T_{\max}$ at which the relative error $\epsilon_R$ in the free energy remains below a certain threshold, $\epsilon_{\mathrm{th}}$, for different values of $g$.
The results, depicted in Fig.~\ref{fig:1d}(b), show that, whereas the method still performs well away from criticality, the range of valid temperatures decreases as the gap grows. 
As we can expect, the range of temperatures decreases when the system size grows, since the effective excitations captured by our ansatz correspond to an $O(1)$ energy excess, and thus to vanishing energy density in the thermodynamic limit. Yet, our results show that the method can be used in combination with finite size ansatzes to explore a sizable low temperature regime in systems of a couple of hundreds sites.
Any function of $\rho$ that is invariant under the $U=\Phi_A \otimes \Phi_B$ isometry can be directly computed in the effective subspace. Remarkably,
our method thus allows computing entanglement quantities that are difficult to access otherwise.

In particular, we benchmark the calculation of entanglement negativity, which
can be used to probe quantum critical points at finite temperature \cite{calabrese2014finite}. 
The difference between the negativity at finite temperature and its value at $T=0$
$\varepsilon_s(T) \equiv \varepsilon(T)-\varepsilon(0)$, 
for a CFT in periodic boundary conditions, is predicted to scale (at intermediate $T$) as
\begin{equation}
     \varepsilon_s(T) = \frac{c}{2} \ln{\bigg(\frac{1 - e^{-2\lambda}}{2\lambda}\bigg)} + f(e^{-2\lambda}),\label{cft}
\end{equation}
where $\lambda = \pi l T$, $l$ is the subsystem size and $c$ is the central charge of the underlying CFT \cite{calabrese2014finite,shapourian2019finite}. In Fig.~\ref{fig:1d}(c), we compare 
the results obtained for $\varepsilon_s(T)$ with our ansatz
with the theoretical prediction. 
By fitting our data to
expression~\eqref{cft}, we obtain a central charge $ c \approx 0.508 $, in excellent agreement with the exact value $ c = 1/2 $. 
At low temperatures, 
the expansion in~\cite{calabrese2014finite} predicts a different behavior for the same quantity.
Our data also fit well this region, as shown by the inset of Fig.~\ref{fig:1d}(c)
(see also Appendix ~\ref{app:negativity}). 
In the opposite limit, at
high temperatures, $\varepsilon_s(T) $ saturates (see Appendix ~\ref{app:cost}), again as expected from CFT predictions ~\cite{calabrese2014finite}.

\begin{figure}
    \centering
     \includegraphics[width=0.75\linewidth]{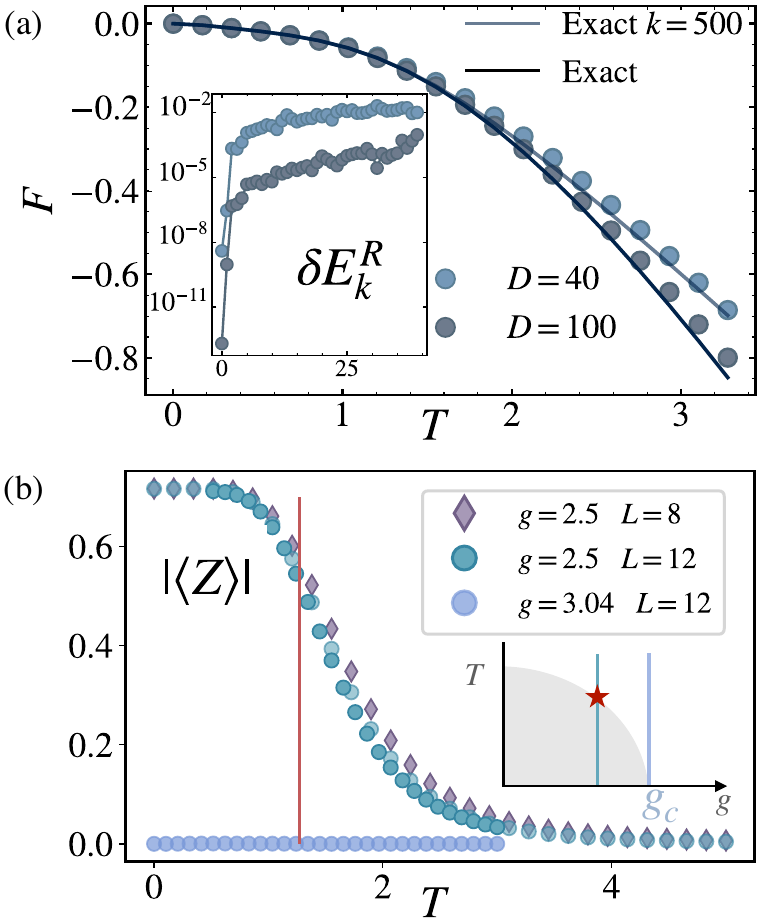}
    \caption{\justifying Numerical results 
    for the 2D TFI model on a $L\times L$ square lattice with periodic boundary conditions.
         (a) Free energy
         (main plot) 
         and relative error in energy spectrum $\delta E_k^R$ (inset)
         at $g=1.8$, $L=4$, for bond dimensions $D=40$ (light circles) and $100$ (dark circles), compared to results from exact diagonalization in full (black solid line), or truncated to a fixed number of excitations (blue solid line). 
       (b) Absolute value of the average magnetization $\abs{\braket{Z}}$
       as a function of temperature. Results are presented at
           $g=2.5$ for $L=8$ (purple diamonds) and $12$ (green circles), and at $g=3.04$ for $L=12$ (blue circles).
           $L=8$ is shown for $D=70$ and $L=12$ is shown for both $D=100$ (darker) and $D=70$ 
           (lighter symbols),
           to illustrate convergence in $D$. 
           The red line indicates the expected critical temperature for $g=2.5$~\cite{czarnik2019finite}, and the 
           inset shows schematically the phase diagram, with the probed values of $g$ marked by vertical lines.} 
    \label{fig:2d}
\end{figure}

\subsection{Numerical results in 2D}
\label{subsec:num2D}

In order to probe the more challenging two-dimensional case, we consider the TFI in Eq.~\eqref{hising} on a square lattice of dimension $L\times L$, for sizes up to $L=12$.
The phase diagram of the model, sketched in the inset of Fig.~\ref{fig:2d} (b), exhibits a quantum phase transition at $g_c\approx 3.044$~\cite{Pfeuty1971}, and a thermal phase transition for $g<g_c$, which occurs at $T_c\approx 2.269$ for $g=0$~\cite{Onsager1944},
 and has been studied in detail with Quantum Monte Carlo simulations~\cite{hesselmann2016thermal}.

To benchmark the method, we first consider a small system $L=4$ at $g=1.8$, for which the energy gap is small at this size.
We compute the energy spectrum $E_k$ and the free energy $F$ using our method, and compare it with exact diagonalization results in Fig.~\ref{fig:2d}(a).
While the first few energy levels are accurately captured, we observe that the relative errors in the eigenenergies $\delta E_k^R$ (shown in the inset) increase rapidly for higher levels. Nevertheless, the free energy is well-approximated in the low-temperature regime, showing that the ability of our ansatz to capture thermodynamic properties goes beyond the ability of reproducing the energy spectrum level by level, consistent with our observations in the 1D case.

We furthermore test the method on larger systems (up to $L=12$) at values of the field $g=2.5$ (for which the transition happens at $T\approx 1.2737$~\cite{hesselmann2016thermal}) and $g=3.04$ (near critical at $T=0$).
Fig.~\ref{fig:2d}(b) shows the temperature dependence of the average magnetization $\abs{\braket{Z}}=\left |\frac{1}{N}\sum_i\sigma_i^z \right |$,
which acts as an order parameter.
At $g=3.04$ we find the order parameter to vanish at all $T$, as the system is in the disordered phase at any $T$. 
Further from the critical point, at $g=2.5$, we observe the magnetization decaying as temperature grows, from a plateau at small $T$ (converged in system size) to a vanishing value at high temperatures, consistent with a disordered phase.
Rather than an abrupt drop, as expected in the thermodynamic limit, we obtain a smooth variation for the finite system, but becoming sharper as $L$ increases, and compatible with the literature~\cite{czarnik2012projected,hesselmann2016thermal,czarnik2019finite,kadow2023isometric}.

\section{Improving the method away from criticality}
\label{sec:non-crit}

As 
intuitively expected from the CFT motivation of our approach, 
and shown in Fig.~\ref{fig:1d}(b),
the performance of the method decreases for gapped systems. For instance, 
for the 1D TFI model studied in~\ref{subsec:num1D}
if we consider the case $g=1.5$, at which the gap is significant, we find the temperature threshold at $T_{\max}\approx 0.03$ for $L=128$,
to be compared to $T_{\max}\approx 0.12$ for the critical $g=1$ at the same size.
In the gapped case, a smaller bond dimension suffices to capture accurately the ground state (and a few excitations), but the corresponding Schmidt vectors
do not capture a large enough subspace \cite{chepiga2017excitation}. 
We may thus try to improve out method by combining it with a multi-excitation ansatz in the spirit of Ref.~\cite{arceci2022entanglement}. 
More concretely, we could find a few of the lowest-energy eigenstates by the recurrent variational optimization, and combine their Boltzmann-weighted individual contributions to the thermal state with that of an effective Hamiltonian obtained from a state different from the ground state. 

\begin{figure}
    \centering
    \includegraphics[width=0.8\linewidth]{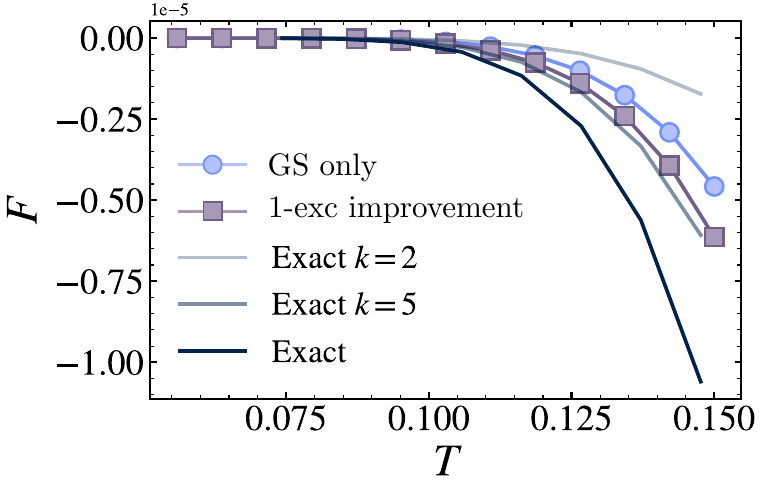}
    \caption{\justifying Performance of our method and its one-excitation extension in the gapped regime.
    The figure shows the free energy as a function of temperature for the $1\mathrm{D}$ TFI at $g=1.5$, for $L=96$, using $D=100$, for our original ansatz (blue circles) and the improved one in \eqref{eq:improved_E1} (squares). Solid lines indicate the result from a fixed number $k$ of exactly computed excitations, from $k=2$ (lightest grey, less accurate than our ansatz), to $k=5$ (comparable to the improved ansatz) and the full exact result (black line, for reference).}
    \label{fig:largegap}
\end{figure}

To test this idea, we variationally find the ground and first excited state of the 1D TFI \eqref{hising}, and write the improved ansatz (up to normalization) as
\begin{equation}
    \rho_{\beta}\propto e^{-{\beta} E_0}\ket{E_0}\bra{E_0}+U_1^\dagger \left ( \sum_{k=1}^{\chi} e^{-\beta \tilde{E}_k^{(1)}}\ket{\tilde{E}_k^{(1)}}\bra{\tilde{E}_k^{(1)}}\right) U_1,
    \label{eq:improved_E1}
\end{equation}
where $U_1$ and $\tilde{E}_k^{(1)}$ are defined in terms of the effective Hamiltonian obtained from the first excitation.
As shown in Fig.~\ref{fig:largegap}, this improves the performance moderately.
The figure compares the results obtained with the original method and with this modification
with exact results. 
First,
we observe that including one additional energy level in~\textbf{}\eqref{eq:improved_E1} 
yields a lower value of the free energy than the version using only the ground state.

To check that the observed improvement does not simply correspond to having two very accurate estimates for the first two energy levels, 
we compare the results with those obtained from keeping only these two levels, but exactly computed (light gray line in Fig.~\ref{fig:largegap}).
This produces a worse value of the free energy than our original method.
Using the modified ansatz~\eqref{eq:improved_E1} seems comparable, instead, to having five exactly computed excitations (darker gray line in the figure), in this case.

By increasing the number of individual excitations, and using the effective Hamiltonian for the highest computed one, 
we could devise a hybrid method that extends the procedure in~\cite{arceci2022entanglement} with an additional effective subspace.

Interestingly, we find that a similarly accurate estimate of the free energy is obtained using only the information from the first excited state, i.e. $\rho_{\beta}\propto U_1^\dagger e^{-\beta \tilde{H}_1} U_1$, where $\tilde{H}_1$ is obtained projecting the Hamiltonian with the Schmidt isometries from the first excited state.
We observe that the spectrum of $\tilde{H}_1$ also captures the ground state energy $E_0$ with good accuracy,
which can be understood since our ansatz gives an approximation for the ground state, and a first excitation orthogonal to this one that will contain some component of the actual ground state.

\section{Discussion}
\label{sec:discussion}
We have presented a method to obtain a low-rank tensor network approximation of the thermal equilibrium state of a quantum many body system at low temperatures, starting from the zero-temperature limit.
Our method is thus complementary to the standard imaginary time evolution approach, which accumulates error as the temperature decreases (see Appendix \ref{app:comparison}).
This approach is inspired by CFT arguments:  as the ground state acts as a primary state from which
to generate excited states through specific local operators, we use
the Schmidt vectors of the ground
state to define a relevant subspace for the approximation of a thermal state. Our numerical results show how the applicability of the method actually extends to more general low-temperature scenarios and higher dimensions. 
While the performance degrades with the ansatz capturing a smaller range of temperatures as the gap opens, it is possible to improve the results by constructing a mixture of our ansatz, computed from an excited state, and lower single excitations components,
in a way interpolating between our method and that in Ref.~\cite{arceci2022entanglement}. 

Our method enables the efficient computation of thermodynamic and entanglement properties. Compared to other multi-excitation methods, we find that our method produces a better approximation with fixed rank at a comparably lower computational effort.
We have presented the algorithm for tree tensor networks, which allows us to study 
finite 2D systems, with different geometries and boundary conditions, despite not inherently fulfilling an area law. 
MPS can straightforwardly be used as a particular case of TTN, but an extension for other TN ansatzes could also be possible. A natural candidate is the multi-scale entanglement renormalization ansatz (MERA) \cite{vidal2008mera}, which effectively captures scale-invariant behavior and supports the construction of excitations~\cite{evenbly2009algorithms}. Applying the method to other two-dimensional TNs like projected entangled pair states (PEPS) is more challenging due to the absence of a canonical form. Although variational excitation methods exist \cite{vanderstraeten2019simulating}, the concrete procedure to apply our method is non-trivial.
There are however subfamilies of PEPS with such canonical structure, as sequentially generated states~\cite{banuls2008sequentially,wei2022sequential}, or isometric TNs~\cite{zalatelisotns},
which could be investigated as a promising platform for extending our method, for computing excitations and thermal states.

In its current version, our method is applicable to systems of finite size, as the ansatz approaches a state with vanishing energy density in the thermodynamic limit. It would be interesting to investigate whether the method can be further extended to describe states with finite energy density in the thermodynamic limit, while preserving some of its computational advantages.

Whereas we focused here on the properties of the ansatz in different settings, practical applications may include the exploration of very-low $T$ thermal phase transitions in 2D systems and comparison with existing 2D tensor networks algorithms at finite temperature, or critical fan regimes~\cite{Frerot2019} in $1\mathrm{D}$.

\begin{acknowledgments}
We thank J. I. Cirac, Y. Liu, T. Roscilde and C. Hotta for inspiring discussions and for pointing us to relevant references. We thank A. Weichselbaum and W. Li for sharing useful data for the comparison of our results with XTRG.
D.C. thanks S. Lu for discussions on 2D models and G. Giudice for feedback on the manuscript.
This work was partially supported by 
the Deutsche Forschungsgemeinschaft (DFG, German Research Foundation) under Germany's Excellence Strategy -- EXC-2111 -- 390814868; and Research Unit FOR 5522 (grant nr. 499180199),  
and by the EU-QUANTERA project TNiSQ (BA 6059/1-1).
\end{acknowledgments}

Raw data are available on Zenodo~\cite{cocchiarella_2024_14336114}.

\appendix

\section{Convergence of the Method}\label{app:convergence}
The convergence of tensor network methods can be evaluated by systematically increasing the bond dimension $D$ of the tensors in the network. As $D$ grows, the solution becomes more accurate, but computationally more expensive. If the Schmidt vectors of the ground state $\ket{\psi_{GS}} = \sum _{\alpha}^{D} \lambda_{\alpha} \ket{\phi_A^{\alpha}}\ket{\phi_B^{\alpha}}$ form a complete basis,  the isometry defined as $U=\Phi_A \otimes \Phi_B$ is a unitary matrix, and our ansatz will converge to the exact result. However, practical convergence is limited by the physical properties of the system under consideration.  At critical points, the Schmidt values typically decay algebraically, allowing for better convergence with increasing $D$ compared to gapped systems, where Schmidt values decay exponentially. 
In this case the ground state may not span the entire Hilbert space, as a finite $D$ can already capture the essential physics of the system. 
Artificially increasing $D$ beyond the rank of the ground state might enlarge the isometry, but the additional singular vectors are no longer derived from the ground state. Consequently, the improvement is not guaranteed to be smooth or systematic as $D$ approaches its exact value. 

\begin{figure}
    \centering
    \includegraphics[width=0.8\linewidth]{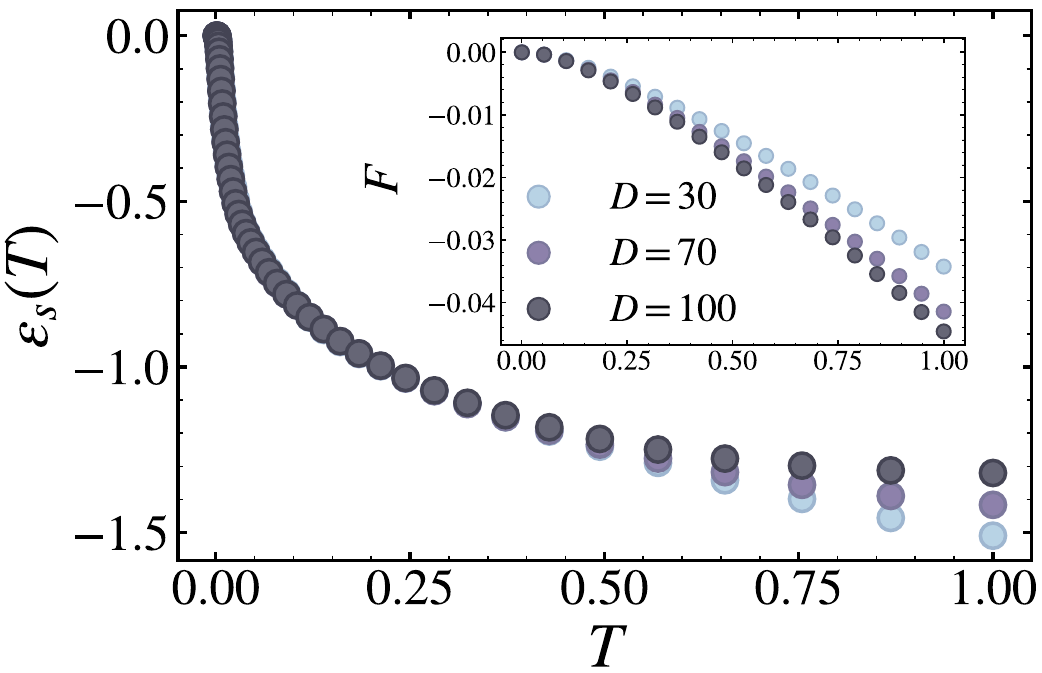}
    \caption{\justifying Convergence of the results with bond dimension as a function of temperature, for the $1\mathrm{D}$ TFI at $g=1,$ $L=128$. The upper panel shows  entanglement negativity in PBC, while
    the main figure shows the free energy in OBC, for the same set of bond dimensions.}
    \label{fig:convergence_negativity}
\end{figure}

In practice, we increase the bond dimension $D$ of the TTN until results are converged within a predefined precision.
Fig.~\ref{fig:convergence_negativity} shows, as a concrete example, the convergence of the entanglement negativity and free energy (inset) with bond dimension for a system of size $L=128,$ at the critical point $g=1$.

In order to evaluate the performance of the method when the exact solution is known, we establish a relative error threshold $\epsilon_\mathrm{th}=0.01$ for the free energy and define $T_{\max}(\epsilon_R<\epsilon_{th})$ as the maximum temperature at which the relative error $\epsilon_R$ in the free energy remains below $\epsilon_{th}$, as shown in Fig.2(b) in the main text.

\section{Computational cost of the algorithm}\label{app:cost}
We summarize the computational costs of the algorithm's key steps, focusing on their scaling with the bond dimension 
$D$. A more detailed understanding of these estimates can be obtained by referring to \cite{schollwock2005density,Verstraete2008} for the MPS case and to \cite{silvi2019tensor} for TTN.

\begin{itemize}
    \item Ground state search and creation of the bond-0 effective Hamiltonian $\tilde{H}$: The ground state search is performed iteratively by optimizing individual tensors in the network. Once the ground state is obtained, the bond-0 effective Hamiltonian $\tilde{H}$ is constructed. The computational cost is dominated by the most expensive tensor contraction in these steps, scaling as $\mathcal{O}(D^3)$ for MPS and as $\mathcal{O}(D^4)$ for TTN.
    
    \item Diagonalization of $\tilde{H}$: $\tilde{H}$ is a $D^2 \times D^2$ matrix. In the current version of the proposed method, we have fully diagonalized $\tilde{H}$, therefore the cost of this step is $\mathcal{O}(D^6)$. 
    If only few energies are needed, the cost can be reduced to $\mathcal{O}(D^4)$. 

\item Entanglement negativity: this step involves two main computational processes. First, the partial transposition of the density matrix, which scales as $\mathcal{O}(D^4)$, and second, the spectrum of the partially transposed matrix, which has a higher computational cost of $\mathcal{O}(D^6)$. 
  \item Expectation values: Computing the expectation value of a local observable acting on a specific site requires the contraction of only three tensors, provided the TTN is center-gauged to the tensor associated with that site. This computation is independent of the system size $N$ and scales as $\mathcal{O}(D^4)$.
\end{itemize}

\section{Relation to other methods}\label{app:comparison}
\begin{figure}
    \centering
    \includegraphics[width=0.8\linewidth]{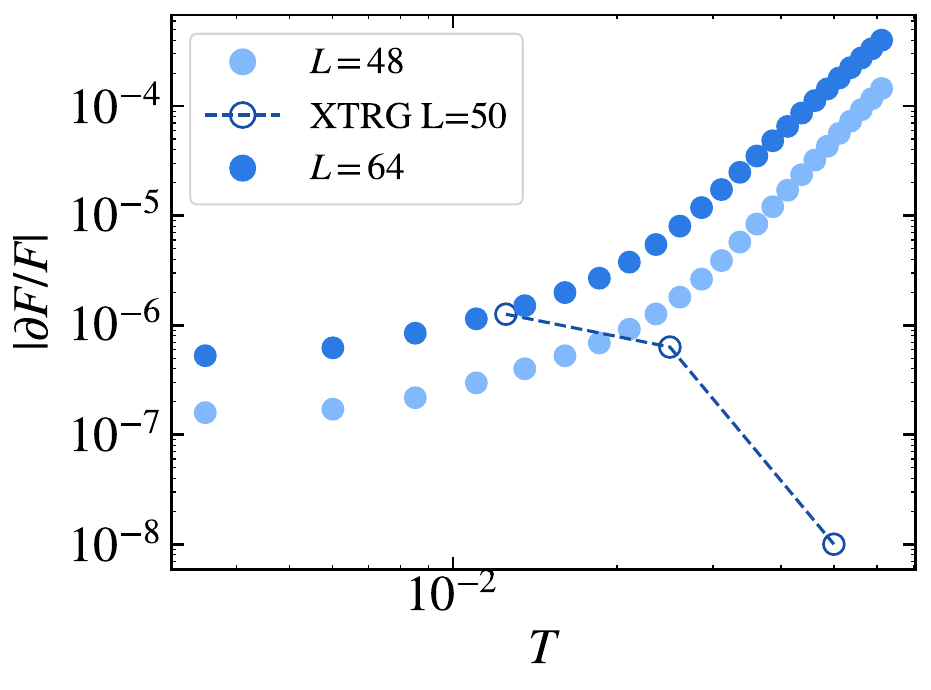}
    \caption{\justifying
    Relative errors of free energy in a spin-$1/2$ XY chain with OBC. The relative error $\abs{\partial F / F}$ is plotted as a function of temperature $T$ for system sizes $L = 48$ (light blue), $L = 64$ (dark blue) using our method, and compared with XTRG data for $L = 50$ (hollow circles), taken from Ref.~\cite{chen2018exponential}. All the data are reproduced with a bond dimension $D=100$.
    }
    \label{fig:comparison}
\end{figure}

Standard imaginary time evolution techniques construct thermal states starting from high temperatures, accumulating errors as the temperature decreases. Our approach, in contrast, targets the low-energy regime directly and it accumulates error as the temperature increases. To illustrate the complementary regimes of applicability between our low-temperature method and standard high-temperature algorithms based on imaginary time evolution, we provide a direct benchmark against published data from Fig. 5 of \cite{chen2018exponential}, where the exponential thermal renormalization group (XTRG) is applied to the spin-1/2 XY chain with open boundary conditions (OBC), for a chain of length $L = 50$.
XTRG achieves excellent accuracy at intermediate to high temperatures, and our approach, tailored for the low-energy regime, maintains a comparable level of accuracy at low-temperatures. This is demonstrated in Fig.\ref{fig:comparison}, where we directly compare the relative error in free energy from our method, for system sizes $L=48,64$ using $D=100$ and those reported in Ref.\cite{chen2018exponential} for $L=50$, using the same bond dimension $D=100$. We take three points corresponding to their lowest simulated temperatures, using their most precise method. The error points emphasizes the natural division of strengths: XTRG is well-suited in the high/intermediate temperature regime, and our method offers an efficient and accurate alternative in the low-temperature limit.

\section{Entanglement negativity at low and high temperature: fit from predictions}\label{app:negativity}
\begin{figure}
    \centering
    \includegraphics[width=0.8\linewidth]{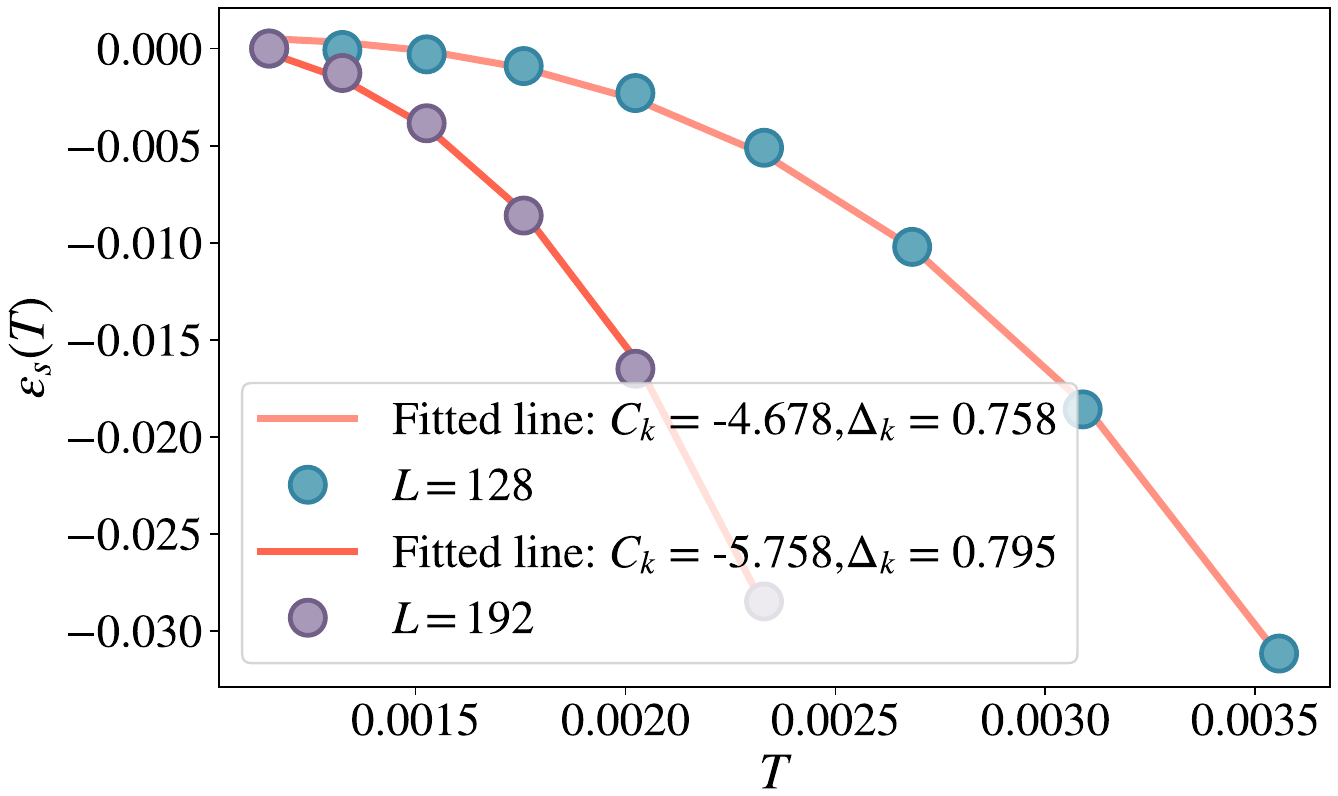}
    \caption{\justifying
    Low-temperature behavior of the entanglement negativity. We show our data for $L=128$ (blue circles) and $L=192$ (purple circles), and the corresponding fits to Eq.~\eqref{eq:lowTneg} (orange lines), for the $1\mathrm{D}$ TFI at criticality.
    }
    \label{fig:fit}
\end{figure}

In the next paragraphs, we provide details about the fits for the entanglement negativity shown in the main text. 
In \cite{calabrese2014finite}, a universal scaling form for the finite temperature negativity was derived. In the limit of low ($T \ll l$, where $T$ is the temperature and $l$ is the subsystem size) and high temperatures ($T \gg l$), the expansion of this universal form can be obtained by means of the operator product expansion (OPE). 
At high-temperature, the negativity attains a constant non-universal value. The study of the approach to this constant value requires the knowledge of the subleading terms in
the OPE, a more involved analytical calculation. It is remarkable that our data finds the saturation naturally, as can be apreciated in Fig.~\ref{fig:convergence_negativity}.
At intermediate $T$, CFT predicts~\cite{calabrese2014finite,shapourian2019finite} the behavior shown in Eq.3 of the main text. Fitting our data to this formula, we indeed recover the central charge value $c=0.508$, in good agreement with the theoretical $c=0.5$.
At low-temperature, the contribution to the negativity is of the form
\begin{equation}
    \varepsilon_s(T)= \frac{c(\pi l T)^2}{12} + C_k (lT)^{4\Delta_k} + \dots.
    \label{eq:lowTneg}
\end{equation}
The first contribution to this expression is analytic and comes from the stress-energy 
tensor. The second, is a non-analytic and 
non-trivial term. The constant $C_k$ is undetermined, and $\Delta_k$ is a scaling dimension~\cite{calabrese2014finite}.
Understanding which contribution is more relevant depends on the precise operator content of the theory and is indeed an interesting check~\cite{calabrese2014finite}. Our method allows us to directly extract these terms numerically, from the fit of
our computed data to the expression in Eq.~\eqref{eq:lowTneg}.
In particular, we obtain for $L=128$, $\Delta_k \sim 0.76$, and for $L=192$, $\Delta_k \sim 0.8$  whilst $C_k\sim -4.6$ for $L=128$, and $C_k\sim -5.7$ for $L=192$, as reported in Fig.~\ref{fig:fit}.

\bibliography{bib}


\end{document}